\documentstyle[preprint,prd,aps,epsfig]{revtex}
\begin{document}
\draft
\preprint{
\vbox{
\halign{&##\hfil\cr
}}
}
\title{The lifetime of $B_c$-meson and some relevant problems}
\author{Chao-Hsi Chang$^{1,2}$,
Shao-Long Chen$^3$, Tai-Fu Feng$^3$ and Xue-Qian Li$^{1,2,3}$}
\address{$^1$  CCAST (World Laboratory), P.O. Box 8730, Beijing 100080,
China}
\address{$^2$  Institute of Theoretical Physics, Academia Sinica, P.O. Box
2735, Beijing 100080, China}
\address{$^3$  Department of Physics, Nankai University, Tianjin 300071,
China}
\maketitle
\begin{center}
\begin{abstract}
The lifetime of the $B_c$-meson is estimated with consistent
considerations on all of the heavy mesons ($B^0, B^\pm, 
B_s, D^0, D^\pm D_s$) and the double heavy meson
$B_c$. In the estimate, the framework,
where the non-spectator effects for nonleptonic
decays are taken into account properly, is 
adopted, and the parameters needed to be fixed
are treated carefully and determined by fitting 
the available data. The bound-state effects in it are
also considered. We find that in decays 
of the meson $B_c$, the QCD correction terms
of the penguin diagrams and 
the main component terms $c_1O_1$, $c_2O_2$
of the effective interaction Lagrangian
have direct interference that causes an enhancement 
about $3\sim 4\%$ in the total width of the $B_c$ meson.
\end{abstract}
\end{center}


\section{Introduction}

\indent

Recently the meson $B_c$ has been observed
by CDF collaboration at Tevatron\cite{Fermi},
that careful studies of the meson $B_c$ 
are motivated with the fresh reason.

It is known that $B_c$ is a meson of the ground state
of a double heavy flavored quark-antiquark system (there is 
no light flavored quark as a valence being involved), and 
its decay must be either through the decays of each component
(a heavy flavor), or through the annihilation of the two 
components (two heavy flavors), that is very different from
$B$ and $D$ mesons. Of the decays, the contribution from 
each component ($\bar{b}$-quark or $c$-quark) 
to the total width happens to be comparable each other
so in the future experiments it is accessible that to 
have precise measurements on each component. 
Namely with the meson of $B_c$ alone, one may 
investigate the two different heavy flavors simultaneously.
Especially, certain decay mechanisms play a similar 
role in $D$-decays, $B$-decays and $B_c$-decays 
so some parameters appearing in $B_c$-decays should be 
the same as those in $B$-decays or in $D$-decays, 
therefore, when one estimates the $B_c$-decay one may
use the experimental available data of $D$-decays 
and $B$-decays as input phenomenologically 
to determine them under the consistent considerations.
Obviously in this way the estimate for $B_c$ 
lifetime should be comparatively reliable. 

The meson $B_c$ certainly is an independent complement
to $B$-mesons and $D$-mesons for studying the two heavy flavor 
$b$ and $c$ decays. Furthermore, if one carries on a comparitive
study of the two heavy flavors, it has unique advantages. 
In this sense, $B_c$-meson will offer an extra 
interesting and unique laboratory for the heavy flavor 
decay studies. 

There have been quite a lot of studies on the lifetime
of $D$ and $B$ mesons and the meson $B_c$ as 
well\cite{Bigi,Neubert,He,Luke,Liuc,Bagan,Cheng,Chang,Bigi1,Bene,Vary,Likh}
The reason in part is that for the lifetime it is comparatively 
`easy'. Due to the duality for quarks and hadrons: 
$$\sum\limits_{i,j} |q_i,g_j\rangle \langle q_i,g_j| = 
\sum\limits_{k} |h_k\rangle \langle h_k|$$ 
where $h_k, q_i$ and $g_j$ denote hadrons, quarks and 
gluons respectively, the optical theorem may be set on 
the level of hadrons or the level of quark-gluons, 
an inclusive processes of hadrons can be turned onto 
a quark level instead. In general, the problem for evaluating 
a decay rate of a hadron is hard, because the relevant hadron 
matrix element cannot be handled reliably. The matrix element 
contains non-perturbative QCD effects, so one cannot compute them 
very satisfactorily based on an 
existent underlying-theory\footnote{In 
principle the lattice gauge simulation may deal with the 
non-perturbative effects as well as one wishes, but in practice 
the computer ability now still is at quite sizable `distance' to 
obtain sufficiently accurate results for calculating such hadron 
matrix elements.}. The optical theorem can be applied for evaluating 
lifetimes and certain inclusive processes, so the problem can be 
`solved' in part: the non-perturbative effects are summed by the 
theorem, generally only in the initial state are still needed be 
handled. Thus in the studies of the lifetime and/or the inclusive
process as well, one may focus the efforts mainly on the decay
mechanisms. 

For the estimate of the lifetimes and inclusive decays, the 
effective Lagrangian for weak decays
with QCD corrections should be known\cite{Bigi,Neubert,He}. 
With the effective Lagrangian for $c$ and $b$ decays, 
phenomenological analyses of $D$ meson lifetimes
and $B$-meson lifetimes have been made\cite{Cheng}.
For all the heavy meson decays
the contributions can be decomposed into
three categories: the dominant one i.e. the direct
decay of the heavy quark while the light quark remains
as a spectator (this contribution is very sensitive
to the heavy quark mass i.e.
proportional to $M_Q^5$); the non-spectator 
one from W-annihilation (or exchange) (WA or WE); 
and the one from the Pauli interference(PI)\cite{Neubert}.
The parameters which are needed to be fixed 
are the quark masses, the matrix element
$\langle 0|J_{\mu 5}|M_{(B_c,B,D)}\rangle$ relating
to the decay constant, and the relevant non-factorizability
parameters etc as well\cite{Neubert}. The range of all
the parameters are known, but their precise values 
are not well calculable. The better way is to fix them
phenomenologically by fitting data. 

To have a better estimate of the lifetime of $B_c$
than before, we will take a `consistent' view of 
the parameters appearing in estimating the lifetimes for 
all of the heavy mesons $D, D_s, B, B_s$
and those in estimating that for the meson $B_c$. 
Namely to estimate the lifetime of the meson $B_c$
with the parameters fixed by phenomenologically
fitting the available 
data for the other heavy mesons. We also try to 
discuss some uncertainties of the estimate in the paper. 

In literatures, the charm quark mass $m_c$ appearing in the 
estimate for D and B
decays takes different values\cite{Cheng,Alek}. 
We think it is reasonable,
because the mass appears in different situations: in the initial
state for D-decays but in the final state for B-decays. In
general, for the quark (antiquark) in the parent meson
of a concerned decay mode, its mass should take
its `pole' value if the bound-state effects are ignored,
whereas, for the masses of the product quarks (antiquarks)
in the final state of an inclusive process, it is more reasonable to
take relevant running masses and the running energy-scale
should be the mass of the decaying quark (or mesons for WA and PI).
Anyhow, this problem is somewhat subtle. In the earlier estimations
for inclusive processes, the quark decays are considered only as
if the quark is `free'. However, some authors have pointed out that
the bound-state effects on the effective mass of the heavy quark
should be taken into account. Namely the heavy quark effective mass,
appearing in the formulation, should deviate from the pole value by 
an amount to correspond to the binding energy\cite{Narison,Kot,wu}. 
In our work, we also pay attention on the effects and use a
parametrization which is a bit different from that of \cite{Kot}
to account for the bound-state effects on the mass of decaying
quark (see the context below for details).

The relation between the pole mass and the running $\overline{MS}$
mass is used many times in our estimation, so we present it up
to one loop level here precisely for convenience. It reads
\begin{equation}
\label{run}
m=\bar m(\bar m)(1+\frac{4}{3}\frac{\alpha_s(\bar m)}{\pi}),
\end{equation}
where the running coupling constant at 1-loop level is
\begin{equation}
\alpha_s=\frac{12\pi}{(33-2n_f)ln\frac{Q^2}{\Lambda^2}},
\end{equation}
with $\alpha_s(m_z^2)=0.118$\cite{Data}. 
The running mass runs as:
\begin{equation}
\bar m(Q^2)=\frac{m}{(\frac{1}{2}ln\frac{Q^2}{\Lambda^2})^{d_m}},
\end{equation}
where $d_m=\frac{12}{33-2n_f}$. 

For the lifetimes of $B$ and $D$ mesons,
the contributions from
penguin terms of the effective Lagrangian
generally are not important\cite{He} because of
smallness of their coefficients $c_3\sim c_6$. 
But as pointed out in \cite{Dai}, the
penguin contributions to the charmless decays of B-mesons
are not negligible. The reason is that for those modes the 
main contributions (since this parts are not zero,
if we return back to the tree level, thus we will call
them as `tree parts' as in the most literature) suffer 
a cancelation $(c_1+c_2/N_c)$ or
$(c_2+c_1/N_c)$, and the `tree part' $c_1O_1+c_2O_2$
does not contribute in addition, thus the penguin 
contributions become important.

As for $B_c$ meson, the problem has not been investigated 
very carefully. For instance, in the earlier paper
\cite{Chang}, the lifetime of $B_c$ was estimated where the bound
state effect was carefully handled in terms of the Bethe-Salpeter
equation, but in the effective Lagrangian the penguin contributions
were ignored. Recently, Beneke and
Buchalla\cite{Bene} also presented an evaluation of the $B_c$
lifetime, where they also ignored the penguins'.
For the spectator mechanism, the contribution from 
the penguin terms in B decays has been estimated by Bagan {\it et
al.}\cite{Bagan}, and their results show that at most it can give
rise to a few thousandths of changes, so in general it can be
neglected. However, for the WA and PI terms, the operators induced by
the penguin diagrams are $\sum\limits_{i=3}^6c_iO_i$ which contain
terms $(\bar s_ib_i)(\bar c_jc_j)$ and
$(\bar s_ib_j)(\bar c_jc_i)$, where $i,j$ are color indices, so
in some $B_c$ decays they can interfere with the `tree part'
$$L^{(tree)}_{eff}=V_{cb}V^*_{cs}[c_1\bar c_L\gamma_\mu b_L\bar s_L
\gamma^\mu c_L+  c_2\bar s_L\gamma_\mu b_L\bar c_L\gamma^\mu c_L].$$
Therefore, in $B-$ and $D-$decays, the contributions relating to
penguins can be proportional only to $|c_i|^2$, or
$|c_i^*c_j|(i,j=3\sim 6)$, so they are small\cite{He}, whereas, in
some $B_c$ decays there may exist
$|c_i^*c_1|$ and $|c_i^*c_2|$ terms, which are not so small, and 
such interference may bring up a few percents of
corrections in its lifetime. Indeed our numerical results 
show the fact that the interference can make a change in 
lifetime of $B_c$ so large as $3\sim 4\%$ of the total.

In the paper, including the penguin contributions and fixing
the parameters with a `consistent' view to fit the data of $B,
B_s$ and $D, D_s$ decays, we re-estimate the lifetimes, the
branching ratios of the semileptonic and pure leptonic decays
of the meson $B_c$. We expect to gain more information on $B_c$
about QCD and decay mechanisms.

The paper is organized as follows: after the introduction, 
we present the formulation in Sec.II, and give the
numerical results and the concerned phenomenological
parameters in Sec.III, then put conclusions and
discussions in the last section. For convenience, we
collect some useful formulars in Appendix A.

\section{Formulation}

\indent

In this section we will describe the 
different mechanisms to the lifetimes for the mesons $D, B$ and 
$B_c$ etc., and present the formulas for later numerical 
calculations. 

\subsection{The spectator components and the contributions
from $b$ or $c$ decays}

With quark-hadron duality and the optical theorem, the `full'
inclusive decay width (the lifetime) of a heavy hadron $H_Q$ 
(containing a heavy quark $Q=b, c$) is related to
the absorptive part of the forward matrix element of the
transition operator $\hat{T}$.
\begin{equation}\label{width}
   {\bf \Gamma}(H_Q\rightarrow X)=\frac{1}{m_{H_Q}}Im \int d^4x
       \langle H_Q|\hat T|H_Q\rangle
   =\frac{1}{2m_{H_Q}}\langle H_Q|\hat \Gamma|H_Q \rangle,
\end{equation}
where
$$
\hat T=T\{i{\cal L}_{eff}(x),{\cal L}_{eff}(0)\},
$$
and ${\cal L}_{eff}$ is the relevant effective weak Lagrangian
which is responsible for the decay. For the concerned final state
$X$ with designated quark-antiquark combination, up-to order $1/m^3_Q$ 
we have:
\begin{eqnarray}
           \nonumber
       \Gamma(H_Q\rightarrow X)&=&\frac{G_F^2m_Q^2}{192\pi^3}|V(CKM)|^2
             {\Big \{}
          c_3^X\langle H_Q|\bar Q Q|H_Q \rangle+
          c_5^X\frac{\langle H_Q|\bar Qi\sigma\cdot GQ|H_Q \rangle}{m^2_Q}
      \\
      & &
        +\sum\limits_ic_{6,i}^X
        \frac{\langle H_Q|(\bar Q\Gamma_iq)(\bar q\Gamma_iQ)|H_Q
\rangle}{m_Q^3}
         +{\cal O}(1/m_Q^4)
             {\Big \}}.
\end{eqnarray}
Here only the heavy quark ($b, c$ quark) decays are 
concerned. In the spectator components of the decays, 
for the heavy meson decays, the light flavor in the heavy meson
remains as a spectator; for the $B_c$-meson decays there are two
possiblities: $\bar {b}$ decays, while
the $c$-quark remains as a spectator,
and $c$ decays, while the $\bar b$-quark remains as a spectator.
In principle, in each spectator components there are two 
`further' components the semileptonic one and the non-leptonic
one:
\begin{eqnarray}
\label{btoc}
\Gamma(b\rightarrow c)&=& \sum_{l=e,\mu,\tau}\Gamma_{b\rightarrow cl\nu}+
\sum_{q=u,d,s,c}\Gamma_{b\rightarrow c\bar qq}
\end{eqnarray}
for $b$-decay;
\begin{eqnarray}
\label{ctos}
\Gamma(c\rightarrow s)&=& \sum_{l=e,\mu}\Gamma_{c\rightarrow s\bar l\nu}+
\sum_{q=u,d,s}\Gamma_{c\rightarrow s\bar qq}
\end{eqnarray}
for $c$-decay. As for the concerned $B_c$ meson, being a double heavy
meson, its two components $\bar b$-quark and $c$-quark, each plays the
decay role and the spectator role once in tern, so both of
eqs.(\ref{btoc},\ref{ctos}) as the spectator components make 
contributions to $B_c$ decay.

The semileptonic and non-leptonic decay rates of
$b-$ quark up-to the order $1/m_b^2$ are evaluated by many authors
\cite{Bigi,Bigi1,Altarelli}.
Since in our numerical computations we need to use their 
formulas, so we quote them from the references into
Appendix A. For $c\to s$, the formulation is similar 
and even simpler, we also include the useful formulas in
Appendix A.

\subsection{The non-spectator components in D and B meson decays}

The non-spectator contributions are crucially important to the $D$ 
inclusive
decays. For instance, the PI contribution may explain the data why
$\tau_{D^{\pm}}\sim 2\tau_{D^0}$, but $\tau_{B^{\pm}}\sim\tau_{B^0}$.
Moreover, the penguin contributions in $D(D_s)$ and $B(B_s)$ decays
are negligible as aforementioned and the bound state effects
emerge. Whereas, all the non-factorization effects 
cannot be reliably well-determined yet.

With straightforward calculations, the precice operators for the
non-spectator contributions may be obtained.

(a) For the $D(D_s)$ decays:

\begin{eqnarray}
            \nonumber
  \Gamma^{WE}(D^0)&=&-\Gamma_0\eta_{nspec}\frac{m_D^2}{m_c^2}
                  (|V_{cs}|^2|V_{ud}|^2+|V_{cd}|^2|V_{us}|^2)(1-x_+)^2
                {\Big \{}
                  (\frac{c_1^2}{N}+2c_1c_2+Nc_2^2)
             \\    \nonumber
              & &
                \times [(1+\frac{x_+}{2})B_1-(1+2x_+)B_2]
                  +2c_1^2[(1+\frac{x_+}{2})\epsilon_1-(1+2x_+)\epsilon_2]
                        {\Big \}}
            \\    \nonumber
              & &
              -\Gamma_0\eta_{nspec}\frac{m_D^2}{m_c^2}
                |V_{cs}|^2|V_{us}|^2\sqrt{1-4x_+} {\Big \{}
                (\frac{c_1^2}{N}+2c_1c_2+Nc_2^2)
             \\   \nonumber
              & &
                \times [(1-x_+)B_1-(1+2x_+)B_2]
                  +2c_1^2[(1-x_+)\epsilon_1-(1+2x_+)\epsilon_2]
               {\Big \}},
       \\        \nonumber
\Gamma^{PI}(D^+)&=&\Gamma_0\eta_{nspec}\frac{p_-^2}{m_c^2}|V_{ud}|^2
                      (|V_{cs}|^2(1-x_-)^2+|V_{cd}|^2)\cdot
                      [(c_1^2+c_2^2)(B_1+6\epsilon_1)+6c_1c_2B_1],
       \\        \nonumber
\Gamma^{WA}(D_s^+)&=&-\Gamma_0\eta_{nspec}\frac{m_{D_s}^2}{m_c^2}
                   |V_{cs}|^2|V_{ud}|^2 {\Big \{}
                   (\frac{c_2^2}{N}+2c_1c_2+Nc_1^2)
                   (B1-B2)+2c_2^2(\epsilon_1-\epsilon_2)
                   {\Big \}}
            \\     \nonumber
            & &
              -\Gamma_0\eta_{nspec}\frac{m_{D_s}^2}{m_c^2}
                   |V_{cs}|^2|V_{us}|^2(1-x_+)^2 {\Big \{}
                   [\frac{c_2^2}{N}+2c_1c_2+Nc_2^2]
             \\     \nonumber
            & &
           \times  [(1+\frac{x_+}{2})B_1-(1+2x_+)B_2]
                   +2c_2^2[(1+\frac{x_+}{2})\epsilon_1-(1+2x_+)\epsilon_2]
          {\Big \}},
       \\  \nonumber
\Gamma^{PI}(D_s^+)&=&\Gamma_0\eta_{nspec}\frac{p_-^2}{m_c^2}|V_{us}|^2
                      (|V_{cs}|^2(1-x_-)^2+|V_{cd}|^2)\cdot
                      [(c_1^2+c_2^2)(B_1+6\epsilon_1)+6c_1c_2B_1],
           \\
\Gamma(D_s^+\rightarrow \tau\nu_{\tau})&=&
                      \frac{G_F^2m_{\tau}^2f_{D_s}^2m_{{D_s}}}{8\pi}
                      |V_{cs}|^2(1-\frac{m^2_{\tau}}{m^2_{D_{s}}})^2.
\end{eqnarray}
where
\begin{eqnarray}
     \nonumber
  \Gamma_0=\frac{G_F^2m_c^5}{192\pi^3},&&
  \eta_{nspec}=16\pi^2\frac{f_{D_q}^2m_{D_q}}{m_c^3},
    \\       \nonumber
  x_+=\frac{\bar m_s^2}{p_+^2};&& p_+=p_c+p_{\bar q},
    \\
  x_-=\frac{\bar m_s^2}{p_-^2};&& p_-=p_c-p_{\bar q}.
\end{eqnarray}
In the equations,
the hadronic parameters are defined as follows:
\begin{eqnarray}
          \nonumber
  \frac{g^{\mu\nu}}{2m_{D_q}}<D_q|O_{\mu\nu}^q|D_q>
                &\equiv& \frac{f^2_{D_q}m_{D_q}}{8}B_1,
       \\
             \nonumber
     \frac{g^{\mu\nu}}{2m_{D_q}}<D_q|T_{\mu\nu}^q|D_q>
              &\equiv& \frac{f^2_{D_q}m_{D_q}}{8}\epsilon_1,
     \\     \nonumber
  \frac{p^\mu p^\nu}{2m_{D_q}^3}<D_q|O_{\mu\nu}^q|D_q>
                &\equiv& \frac{f^2_{D_q}m_{D_q}}{8}B_2\; ,
       \\
   \frac{p^\mu p^\nu}{2m_{D_q}^3}<D_q|T_{\mu\nu}^q|D_q>
            &\equiv& \frac{f^2_{D_q}m_{D_q}}{8}\epsilon_2\; ,
\end{eqnarray}
where
\begin{eqnarray}
     \nonumber
     O_{\mu\nu}^q&=&\bar c\gamma_\mu Lq\bar q\gamma_\nu Lc,
     \\
     T_{\mu\nu}^q&=&\bar c\gamma_\mu T^a Lq\bar q\gamma_\nu T^a Lc\; ,
\end{eqnarray}
with $T^a=\frac{\lambda^a}{2}$ and $\lambda^a$ being the Gell-Mann Matrices.

(b) For the $B(B_s)$ decays:

\begin{eqnarray}
    \nonumber
\Gamma^{WE}(B_d^0)&=&-\Gamma_0\eta_{nspec}|V_{ud}|^2(1-z_+)^2{\Big \{}
              (\frac{c_1^2}{N}+2c_1c_2+Nc_2^2)[(1+\frac{z_+}{2})B_1-(1+2z_+)B_2]
            \\  \nonumber
             & &
              +2c_1^2[[(1+\frac{z_+}{2})\epsilon_1-(1+2z_+)\epsilon_2]]
      {\Big \}}
       \\   \nonumber
       & &
           -\Gamma_0\eta_{nspec}|V_{cd}|^2\sqrt{1-4z_+}{\Big \{}
           (\frac{c_1^2}{N}+2c_1c_2+Nc_2^2)[(1-z_+)B_1-(1+2z_+)B_2]
        \\ & &
        \nonumber
           +2c_1^2[[(1-z_+)\epsilon_1-(1+2z_+)\epsilon_2]]
       {\Big \}},
       \\
       \nonumber
\Gamma^{PI}(B^-)&=&\Gamma_0\eta_{nspec}\frac{p_-^2}{m_B^2}(1-z_-)^2
                [(c_1^2+c_2^2)(B_1+6\epsilon_1)+6c_1c_2B_1],
     \\
      \nonumber
\Gamma^{WE}(B_s^0)&=&-\Gamma_0\eta_{nspec}|V_{us}|^2(1-z_+)^2{\Big \{}
               (\frac{c_1^2}{N}+2c_1c_2+Nc_2^2)[(1+\frac{z_+}{2})B_1-(1+2z_+)B_2]
             \\  \nonumber
              & &
               +2c_1^2[[(1+\frac{z_+}{2})\epsilon_1-(1+2z_+)\epsilon_2]]
       {\Big \}}
        \\   \nonumber
        & &
            -\Gamma_0\eta_{nspec}|V_{cs}|^2\sqrt{1-4z_+}{\Big \{}
            (\frac{c_1^2}{N}+2c_1c_2+Nc_2^2)[(1-z_+)B_1-(1+2z_+)B_2]
         \\ & &
            +2c_1^2[[(1-z_+)\epsilon_1-(1+2z_+)\epsilon_2]]
        {\Big \}}.
\end{eqnarray}
where
\begin{eqnarray}
    \nonumber
    \Gamma_0=\frac{G_F^2m_b^5}{192\pi^3}|V_{cb}|^2,
    &&
    \eta_{nspec}=16\pi^2\frac{f_{B_q}^2m_{B_q}^3}{m_b^5},
    \\
    z_+=\frac{\bar m_c}{m_{B_q}^2},
    &&
    z_-=\frac{\bar m_c^2}{p_-^2}=\frac{\bar m_c^2}{(p_b-p_{\bar u})^2}.
\end{eqnarray}
Analogous to D meson, the parameters $B_1, B_2, \epsilon_1$ and
$\epsilon_2$
are defined:
\begin{eqnarray}
          \nonumber
  \frac{g^{\mu\nu}}{2m_{B_q}}<B_q|O_{\mu\nu}^q|B_q>
                &\equiv& \frac{f^2_{B_q}m_{B_q}}{8}B_1,
       \\
             \nonumber
     \frac{g^{\mu\nu}}{2m_{B_q}}<B_q|T_{\mu\nu}^q|B_q>
              &\equiv& \frac{f^2_{B_q}m_{B_q}}{8}\epsilon_1,
     \\     \nonumber
  \frac{p^\mu p^\nu}{2m_{B_q}^3}<B_q|O_{\mu\nu}^q|B_q>
                &\equiv& \frac{f^2_{B_q}m_{B_q}}{8}B_2,
       \\
  \frac{p^\mu p^\nu}{2m_{B_q}^3}<B_q|T_{\mu\nu}^q|B_q>
            &\equiv& \frac{f^2_{B_q}m_{B_q}}{8}\epsilon_2.
\end{eqnarray}
where
\begin{eqnarray}
     \nonumber
     O_{\mu\nu}^q&=&\bar b\gamma_\mu Lq\bar q\gamma_\nu Lb\; ,
     \\
     T_{\mu\nu}^q&=&\bar b\gamma_\mu T^a Lq\bar q\gamma_\nu T^a Lb\; .
\end{eqnarray}

\subsection{The non-spectator components in $B_c$ decays}

As pointed above, the spectator contribution to the $B_c$
lifetime is a sum of that from $\bar b$ and
$c$ individual decays as pointed above:
\begin{equation}
\Gamma^{spectator}=\Gamma^{spectator}_b+\Gamma^{spectator}_{\bar c}\; ,
\end{equation}
and $\Gamma^{spectator}_b$ and $\Gamma^{spectator}_{\bar c}$
are the same as they are in $B$ and $D$ decays and given 
in eqs.(\ref{btoc},\ref{ctos}).
Now let us deal with the non-spectator contributions which are
different from that in $B$ and $D$ decays.

To estimate the non-spectator
components in the $B_c$ decays, let us write 
the relevant effective Lagrangian 
precisely here:
\begin{eqnarray}
 L_{eff}^{\Delta C=1}(\mu=m_c)&=&-\frac{4G_F}{\sqrt2}V_{cs}V_{ud}^\ast
       {\Big \{}
       c_1(\mu)(\bar s\gamma_\mu Lc)(\bar u\gamma^\mu Ld)+
       c_2(\mu)(\bar u\gamma_\mu Lc)(\bar s\gamma^\mu Ld)
        {\Big \}}+h.c.,
\end{eqnarray}
and
\begin{eqnarray}
      \nonumber
 L_{eff}^{\Delta B=1}(\mu=m_b)&=& -\frac{4G_F}{\sqrt2} {\Big \{}
   V_{cb}[V_{ud}^*(c_1(\mu)O_1^u+c_2(\mu)O_2^u)+
           V_{cs}^*(c_1(\mu)O_1^c+c_2(\mu)O_2^c)+
  \\
  & &
  \sum\limits_{l=e,\tau,\mu}\bar l\gamma_\mu L\nu\bar c\gamma^\nu L
  b+V_{cs}^\ast\sum\limits_{i=3}^6 c_iO_i]
  {\Big \}}+h.c.\;,
\end{eqnarray}
where the operators are
\begin{eqnarray}
           \nonumber
     O_1^c&=&\bar s\gamma_\mu Lc\bar c\gamma^\mu Lb,
 \\    \nonumber
     O_1^u&=&\bar d\gamma_\mu Lu\bar c\gamma^\mu Lb,
 \\    \nonumber
     O_2^c&=&\bar s_i\gamma_\mu Lc_j\bar c_j\gamma^\mu Lb_i,
 \\    \nonumber
     O_2^u&=&\bar d_i\gamma_\mu Lu_j\bar c_j\gamma^\mu Lb_i,
 \\   \nonumber
     O_3&=&\bar s\gamma_\mu Lb\bar c\gamma^\mu Lc,
 \\   \nonumber
     O_4&=&\bar s_i\gamma_\mu Lb_j\bar c_j\gamma^\mu Lc_i,
 \\   \nonumber
     O_5&=&\bar s\gamma_\mu Lb\bar c\gamma^\mu Rc,
 \\
     O_6&=&\bar s_i\gamma_\mu Lb_j\bar c_j\gamma^\mu Rc_i\; ,
\end{eqnarray}
and $c_i (i=1,2,\cdots)$, denoting the Wilson coefficients due to QCD
corrections, will take the values as those in ref.\cite{He}.
Here we consider the non-spectator components
in $B_c$ decays by two steps. The first step is to compute the relevant
operators upto the order $O(1/m^4_Q)$ and then to evaluate the
contributions precisely.

\subsubsection{Pauli interference (PI) operators}

The Pauli interference (PI) operators $\hat{\Gamma}^{PI}_{tree}$ and
$\hat{\Gamma}^{PI}_{penguin}$
which correspond to the non-leptonic decay induced by
the tree part and penguin respectively are given by:

\begin{eqnarray}
         \nonumber
\hat\Gamma^{PI}_{tree}&= 
&\frac{2G_F^2}{\pi}|V_{cb}|^2|V_{cs}|^2(1-z_-)^2p_-^2
                      \cdot {\Big \{}
                    2c_1c_2\cdot \bar b^i\gamma_\mu Lc^i\bar
c^j\gamma^\mu Lb^j
                    +(c_1^2+c_2^2)
            \\      \nonumber
            & &
                  \cdot
                  \bar b^i\gamma_\mu Lc^j\bar c^j\gamma^\mu Lb^i
                    {\Big \}},
       \\    \nonumber
\hat\Gamma^{PI}_{penguin}& 
=&\frac{2G_F^2}{\pi}|V_{cb}|^2|V_{cs}|^2(1-z_-)^2p_-^2
                     \cdot {\Big \{}
                    (2c_1c_3+2c_2c_4+2c_3c_4)
                    \cdot\bar b^i\gamma_\mu Lc^i\bar c^j\gamma^\mu Lb^j
            \\      \nonumber
           & &
                  +(c_3^2+c_4^2+2c_1c_4+2c_2c_3)\cdot
                    \bar b^i\gamma_\mu Lc^j\bar c^j\gamma^\mu Lb^i
                     {\Big \}}
            \\     \nonumber
            & &
               +\frac{G_F^2}{3\pi}|V_{cb}|^2|V_{cs}|^2(1-z_-)^2\cdot
                  {\Big \{}
                  (1-z_-)p_-^2g^{\mu\nu}+2(1+2z_-)p_-^\mu p_-^\nu
                  {\Big \}}
            \\    \nonumber
            & &
                  \cdot{\Big \{}
                  2c_5c_6\cdot\bar b^i\gamma_\mu Lb^j\bar c^j\gamma_\nu Rc^i
                  +(c_5^2+c_6^2)\cdot\bar b^i\gamma_\mu Lb^i\bar
c^j\gamma_\nu Rc^j
                  {\Big \}}
            \\       \nonumber
            & &
             -\frac{G_F^2}{\pi}|V_{cb}|^2|V_{cs}|^2(1-z_-)^2\bar m_cp^\alpha
                    \cdot {\Big \{}
                      [c_2c_6+c_3c_6+c_1c_5+c_4c_5]
            \\         \nonumber
            & &
                       \cdot
                       [\bar b^i\gamma^\mu Lc^i\bar
c^j\gamma_\alpha\gamma_\mu Lb^j
                       +\bar b^i\gamma_\mu\gamma_\alpha Rc^i\bar
c^j\gamma^\mu Lb^j]
                       +[c_2c_5+c_3c_5+c_1c_6+c_4c_6]
           \\
             & &
                       \cdot
                       [\bar b^i\gamma^\mu Lc^j\bar
c^j\gamma_\alpha\gamma_\mu Lb^i
                       +\bar b^i\gamma_\mu\gamma_\alpha Rc^j\bar
c^j\gamma^\mu Lb^i]
                    {\Big \}}\; ,
\end{eqnarray}
where
\begin{eqnarray}
      z_-=\frac{\bar m_c^2}{p_-^2}, &&
      p_-=p_b-p_{\bar c}\;.
\end{eqnarray}

\subsubsection{Weak annihilation (WA) operators}
The weak annihilation operators are
$\hat{\Gamma}^{WA}_{tree}$,
$\hat{\Gamma}^{WA}_{penguin}$ and
$\hat{\Gamma}^{WA}(B_c\rightarrow\tau\nu_{\tau})$
which correspond to the non-leptonic decay induced by
the tree part, penguin and the pure leptonic (PL) decay 
respectively.\footnote{Due to hilicity suppression, the
decays $B_c\to l(e,\mu)+\nu$ are ignorable for the low
order estimate of the lifetime, thus we do so here.}

\begin{eqnarray}
          \nonumber
\hat\Gamma^{WA}_{tree}& 
=&-\frac{2G_F^2}{3\pi}|V_{cb}|^2|V_{cs}|^2(1-z_+)^2
                 \cdot       {\Big \{}
                    (1+\frac{z_+}{2})p_+^2g^{\mu\nu}-(1+2z_+)p_+^\mu p_+^\nu
                   {\Big \}}
            \\
            & &
                 \cdot{\Big \{}
               (Nc_1^2+2c_1c_2)\cdot\bar b^i\gamma_\mu Lc^i\bar
c^j\gamma_\nu Lb^j
               +c_2^2\cdot\bar b^i\gamma_\mu Lc^j\bar c^j\gamma_\nu Lb^i
                  {\Big \}},
          \\          \nonumber
\hat\Gamma^{WA}_{penguin}&=&-\frac{2G_F^2}{3\pi}|V_{cb}|^2|V_{cs}|^2(1-z_+)^2
                     \cdot    {\Big \{}
                    (1+\frac{z_+}{2})p_+^2g^{\mu\nu}-(1+2z_+)p_+^\mu p_+^\nu
                   {\Big \}}
            \\    \nonumber
            & &
                 \cdot{\Big \{}
               (Nc_4^2+2Nc_1c_4+2c_3c_4+2c_1c_3+2c_2c_4)
                    \cdot\bar b^i\gamma_\mu Lc^i\bar c^j\gamma_\nu Lb^j
            \\   \nonumber
            & &
               +(c_3^2+2c_2c_3)\cdot\bar b^i\gamma_\mu Lc^j\bar
c^j\gamma_\nu Lb^i
                  {\Big \}}
            \\      \nonumber
            & &
               +\frac{4G_F^2}{\pi}|V_{cb}|^2|V_{cs}|^2(1-z_+)^2p_+^2\cdot
                    [(Nc_6^2+2c_5c_6)\cdot\bar b^iRc^i\bar c^jLb^j
                    +c_5^2\cdot\bar b^iRc^j\bar c^jLb^i]
           \\     \nonumber
            & &
            +\frac{2G_F^2}{\pi}|V_{cb}|^2|V_{cs}|^2(1-z_+)^2\bar m_cp_+^\mu
                     \cdot {\Big\{}
               [Nc_6(c_1+c_4)+c_5(c_1+c_4)
           \\     \nonumber
           & &
                +c_6(c_2+c_3)]\cdot[\bar b^iRc^i\bar c^j\gamma_\mu Lb^j
                       +\bar b^i\gamma_\mu Lc^i\bar c^jLb^j]
         \\
         & &
                +(c_2c_5+c_3c_5)\cdot[\bar b^iRc^j\bar c^j\gamma_\mu Lb^i
                   +\bar b^i\gamma_\mu Lc^j\bar c^jLb^i]
                            {\Big\}},
        \\
\hat\Gamma^{WA}(B_c\rightarrow \tau\nu_\tau)
                     &=&
                     -\frac{2G_F^2}{3\pi}|V_{cb}|^2(1-z_\tau)^2\cdot
                           {\Big \{}
             (1+ \frac{z_\tau}{2})p_+^2g^{\mu\nu}-(1+2z_\tau)p_+^\mu p_+^\nu
                           {\Big \}}
              \\
              & &
           \cdot\bar b^i\gamma_\mu Lc^i\bar c^j\gamma_\nu Lb^j\;,
\end{eqnarray}
where the parameters $p_+$, $z_+$ and $z_\tau$ are defined by
\begin{eqnarray}
      \nonumber
    p_+ &=& p_b+p_c,
   \\   \nonumber
    z_+ &=& \frac{\bar m_c^2}{p_+^2}=\frac{\bar m_c^2}{M_{B_c}^2},
    \\
    z_\tau &=& \frac{m_\tau^2}{p_+^2}=\frac{m_\tau^2}{M_{B_c}^2}\;.
\end{eqnarray}

\subsubsection{The contributions from the non-spectator WA and PI 
to the lifetime for $B_c$ meson}

With the optical theorem, substituting all the above 
operators $\hat\Gamma^{WA},\;\hat\Gamma^{PI}$ into the relevant 
matrix element, we may estimate the non-spectator contributions 
to the lifetime of $B_c$ meson:
\begin{equation}\label{bcwidth}
    \Gamma=\frac{1}{2M_{B_c}}\langle Bc|\hat\Gamma|Bc\rangle\;,
\end{equation}
where $\hat\Gamma$ denotes the relevant operators for $PI$ and $WA$
given in the above subsections.

According to eq.(\ref{bcwidth} to evaluate the lifetime, finally 
some hadronic matrix elements need to be determined, whereas,
having nonpertubative nature, they cannot be determined by 
well-established theories so far. Let us discuss their deturmination
here for a while.

First of all, some parameters, such as $B-1, B_2,
\tilde B_1, \tilde B_2, \epsilon_1, \epsilon_2, \tilde\epsilon_1$
and $\tilde\epsilon_2$, appear in 
the corresponding estimates for $B$ and $D$ decays too.
Precisely for $B_c$ decays, they are
\begin{eqnarray}
      \nonumber
    \frac{1}{2M_{B_c}}\langle Bc|O_{V-A}^c|Bc\rangle
              &\equiv& \frac{f_{Bc}^2M_{B_c}}{8}B_1,
         \\     \nonumber
    \frac{1}{2M_{B_c}}\langle Bc|O_{S-P}^c|Bc\rangle
              &\equiv& \frac{f_{Bc}^2M_{B_c}}{8}B_2,
         \\       \nonumber
    \frac{1}{2M_{B_c}}\langle Bc|T_{V-A}^c|Bc\rangle
               &\equiv& \frac{f_{Bc}^2M_{B_c}}{8}\epsilon_1,
         \\          \nonumber
    \frac{1}{2M_{B_c}}\langle Bc|T_{S-P}^c|Bc\rangle
              &\equiv& \frac{f_{Bc}^2M_{B_c}}{8}\epsilon_2,
         \\         \nonumber
    \frac{1}{2M_{B_c}}\langle Bc|\tilde O_{V-A}^c|Bc\rangle
              &\equiv& \frac{f_{Bc}^2M_{B_c}}{8}\tilde B_1,
          \\         \nonumber
     \frac{1}{2M_{B_c}}\langle Bc|\tilde O_{S-P}^c|Bc\rangle
              &\equiv& \frac{f_{Bc}^2M_{B_c}}{8}\tilde B_2,
             \\      \nonumber
     \frac{1}{2M_{B_c}}\langle Bc|\tilde T_{V-A}^c|Bc\rangle
               &\equiv& \frac{f_{Bc}^2M_{B_c}}{8}\tilde \epsilon_1,
             \\      \nonumber
     \frac{1}{2M_{B_c}}\langle Bc|\tilde T_{S-P}^c|Bc\rangle
               &\equiv& \frac{f_{Bc}^2M_{B_c}}{8}\tilde \epsilon_2,
\end{eqnarray}
where the relevant four-quark operators are
\begin{eqnarray}
           \nonumber
       O_{V-A}^c&=&\bar b\gamma_\mu Lc\bar c\gamma^\mu Lb,
       \\     \nonumber
       O_{S-P}^c&=&\bar bLc\bar cRb,
      \\      \nonumber
       T_{V-A}^c&=&\bar b\gamma_\mu LT^ac\bar c\gamma^\mu LT^ab,
      \\       \nonumber
       T_{S-P}^c&=&\bar bLT^ac\bar cRT^ab,
      \\       \nonumber
    \tilde O_{V-A}^c&=&\bar b\gamma_\mu Rc\bar c\gamma^\mu Rb,
       \\      \nonumber
    \tilde O_{S-P}^c&=&\bar bRc\bar cLb,
      \\        \nonumber
    \tilde T_{V-A}^c&=&\bar b\gamma_\mu RT^ac\bar c\gamma^\mu RT^ab,
      \\
    \tilde T_{S-P}^c&=&\bar bRT^ac\bar cLT^ab\;.
\end{eqnarray}
There are eight extra matrix elements corresponding to the newly emerged operators
in the $B_c$ case. The `new' matrix elements relate to the above parameters or
new ones ($\epsilon_3, \epsilon_4, \epsilon_5, \epsilon_6$) as follows:
\begin{eqnarray}
      \nonumber
     \frac{1}{2M_{B_c}}\langle Bc|\bar b Lc\bar cLb|Bc\rangle
         &\equiv& \frac{-f_{Bc}^2M_{B_c}}{8} B_2,
       \\          \nonumber
     \frac{1}{2M_{B_c}}\langle Bc|\bar b Rc\bar cRb|Bc\rangle
                &\equiv& \frac{-f_{Bc}^2M_{B_c}}{8} B_2,
        \\        \nonumber
     \frac{1}{2M_{B_c}}\langle Bc|\bar b\gamma_\mu Lc\bar c\gamma^\mu
Rb|Bc\rangle
                &\equiv& \frac{-f_{Bc}^2M_{B_c}}{8} B_1,
        \\       \nonumber
     \frac{1}{2M_{B_c}}\langle Bc|\bar b\gamma_\mu Rc\bar c\gamma^\mu
Lb|Bc\rangle
                &\equiv& \frac{-f_{Bc}^2M_{B_c}}{8} B_1,
       \\       \nonumber
      \frac{1}{2M_{B_c}}\langle Bc|\bar b LT^ac\bar cLT^ab|Bc\rangle
                 &\equiv& \frac{-f_{Bc}^2M_{B_c}}{8}\epsilon_3,
       \\       \nonumber
     \frac{1}{2M_{B_c}}\langle Bc|\bar b RT^ac\bar cRT^ab|Bc\rangle
                   &\equiv& \frac{-f_{Bc}^2M_{B_c}}{8}\epsilon_4,
       \\       \nonumber
     \frac{1}{2M_{B_c}}\langle Bc|\bar b\gamma_\mu LT^ac\bar c\gamma^\mu
RT^ab|Bc\rangle
                 &\equiv& \frac{-f_{Bc}^2M_{B_c}}{8}\epsilon_5,
             \\
    \frac{1}{2M_{B_c}}\langle Bc|\bar b \gamma_\mu RT^ac\bar c\gamma^\mu
LT^ab|Bc\rangle
                  &\equiv& \frac{-f_{Bc}^2M_{B_c}}{8}\epsilon_6.
\end{eqnarray}

Generally speaking, we may assume that
\begin{equation}
\label{assup}
\tilde B_{1(2)}=B_{1(2)};\; \hspace*{0.2cm} \tilde \epsilon_{1(2)}=\epsilon_{1(2)},
\end{equation}
with symmetry consideration. As for the parameters $\epsilon_{3,4}$ and
$\epsilon_{5,6}$, we would conjecture that
$\epsilon_{3,4}\simeq \epsilon_2$ and $\epsilon_{5,6}\simeq \epsilon_1$
instead of precise computation\footnote{The conjecture
should be tested and proved later on. It should be considered
as a working assumption. With the assumption eq.(\ref{assup}) in addition, 
all the assuptions here will cause an essential uncertainty for the 
estimates.}.

In the earlier literatures, usually $B_1\approx B_2\sim 1$ and
$\epsilon_1\sim -0.15$ from the lattice calculations and $\epsilon_2=0$.
According to our numerical computations and 
trials to fit the data about the lifetimes of
the heavy mesons $D^{\pm}, D^0, D_s, B^{\pm}, B^0$ and $B_s$
and their semileptonic decay branching ratios as well, we 
find that to adjust the values of them and the pole
masses of $b$ and $c$ quarks, when the parameter $\epsilon_2\neq
0$ etc, a better fit is obtained.

Now for the nonspectator component $PI$, we have
\begin{eqnarray}
{\bf \Gamma}^{PI}_{tree}&=&
         \frac{G_F^2}{4\pi}f_{Bc}^2M_{B_c}|V_{cb}|^2|V_{cs}|^2(1-z_-)^2p_-^2
            \cdot  {\Big \{}
           [2c_1c_2+\frac{1}{N}(c_1^2+c_2^2)]B_1
        \nonumber \\
       && +2(c_1^2+c_2^2)\epsilon_1
         {\Big \}}
    \\
{\bf\Gamma}^{PI}_{penguin}&=&
        \frac{G_F^2}{4\pi}f_{Bc}^2M_{B_c}|V_{cb}|^2|V_{cs}|^2(1-z_-)^2p_-^2
          \cdot        {\Big \{}
            [2c_2c_4+2c_1c_3+2c_3c_4+
       \\    \nonumber
       & &
             \ \    \frac{1}{N}(c_3^2+c_4^2+2c_2c_3+2c_1c_4)]B_1
                 +2(c_3^2+c_4^2+2c_2c_3+2c_1c_4)\epsilon_1
        {\Big \}}
   \\     \nonumber
   & &
    -\frac{G_F^2}{4\pi}f_{Bc}^2M_{B_c}|V_{cb}|^2|V_{cs}|^2(1-z_-)^2
          \cdot     {\Big \{}
          [2c_5c_6+\frac{1}{N}(c_5^2+c_6^2)][\frac{2+z_-}{3}p_-^2\tilde B_2
   \\      \nonumber
   & &
        -\frac{1+2z_-}{6}(m_b^2\tilde B_1+m_c^2B_1-4m_bm_cB_2+2m_bm_cB_1)]
           +2(c_5^2+c_6^2)[\frac{2+z_-}{3}p_-^2\tilde\epsilon_2
   \\        \nonumber
   & &
             -\frac{1+2z_-}{6}(m_b^2\tilde\epsilon_1
             +m_c^2\epsilon_1-2m_bm_c(\epsilon_3+\epsilon_4)
              +m_bm_c(\epsilon_5+\epsilon_6))]
           {\Big \}}
    \\    \nonumber
   & & 
     -\frac{G_F^2}{8\pi}
f^2_{Bc}M_{B_c}|V_{cb}|^2|V_{cs}|^2(1-z_-)^2\bar m_c
       \cdot    {\Big \{}
           [c_1c_5+c_2c_6+c_3c_6+c_4c_5
     \\   \nonumber
     & &
           +\frac{1}{N}(c_1c_6+c_2c_5+c_4c_6+c_3c_5)][2m_cB_1
           +m_b(-4B_2+2B_1)]
       \\
      & &
           +2(c_1c_6+c_2c_5+c_4c_6+c_3c_5) 
[2m_c\epsilon_1-2m_b(\epsilon_3+\epsilon_4)
                +m_b(\epsilon_5+\epsilon_6)]
             {\Big \}}\;,
\end{eqnarray}
and for $WA$, we have
\begin{eqnarray}
   {\bf\Gamma}^{WA}_{tree}&=&
        -\frac{G_F^2}{12\pi}|V_{cb}|^2|V_{cs}|^2f_{Bc}^2M_{B_c}(1-z_+)^2
        \cdot      {\Big \{}
              [Nc_1^2+2c_1c_2+\frac{c_2^2}{N}]
     \\   \nonumber
       & &
                \times [(1+\frac{z_+}{2})M_{B_c}^2 B_1
                      -(1+2z_+)(m_b^2B_2+m_c^2\tilde B_2+2m_bm_c B_2)]
       \\   \nonumber
       & &
            +2c_2^2[(1+\frac{z_+}{2})M_{B_c}^2 \epsilon_1
                     -(1+2z_+)(m_b^2\epsilon_2+m_c^2\tilde \epsilon_2
                     +m_bm_c(\epsilon_3+\epsilon_4))]
           {\Big \}},     
\end{eqnarray}
\begin{eqnarray}
{\bf\Gamma}^{WA}(B_c\rightarrow \tau\nu)&=&
          -\frac{G_F^2}{12\pi}|V_{cb}|^2|V_{cs}|^2f_{Bc}^2M_{B_c}(1-z_+)^2
          \cdot      {\Big \{}
               (1+\frac{z_\tau}{2})M_{B_c}^2 B_1       \\ \nonumber
       & &
               -(1+2z_\tau)(m_b^2B_2+m_c^2\tilde B_2+2m_bm_c B_2)
           {\Big \}},
\end{eqnarray}
\begin{eqnarray}
{\bf\Gamma}^{WA}_{penguin}
         &=&
           -\frac{G_F^2}{12\pi}|V_{cb}|^2|V_{cs}|^2f_{B_c}^2M_{B_c}(1-z_+)^2
          \cdot    {\Big \{}
                [(\frac{2c_2+c_3}{N}+2c_1+c_4)(c_3+Nc_4)]
           \\       \nonumber
           & &
                \times [(1+\frac{z_+}{2})M_{B_c}^2 B_1
                 -(1+2z_+)(m_b^2B_2+m_c^2\tilde B_2+2m_bm_c B_2)]
       \\       \nonumber
       & &
             +2(2c_2+c_3)c_3\cdot [(1+\frac{z_+}{2})p_+^2\epsilon_1
                  -(1+2z_+)(m_b^2\epsilon_2+m_c^2\tilde \epsilon_2
                  +m_bm_c(\epsilon_3+\epsilon_4))]
          {\Big \}}
       \\        \nonumber
        & &
           +\frac{G_F^2}{2\pi}|V_{cb}|^2|V_{cs}|^2f_{Bc}^2M_{B_c}^3(1-z_+)^2
          \cdot    {\Big \{}
                   [\frac{c_5^2}{N}+2c_5c_6+Nc_6^2]\tilde B_2
                   +2c_5^2\tilde\epsilon_2
           {\Big \}}
       \\     \nonumber
       & &
          -\frac{G_F^2}{4\pi}|V_{cb}|^2|V_{cs}|^2f_{Bc}^2M_{B_c}\bar
m_c(1-z_+)^2
            \cdot     {\Big \{}
               [(\frac{c_2+c_3}{N}+c_1+c_4)(c_5+Nc_6)]
      \\  \nonumber
      & &
             \times [2m_b B_2+2m_c\tilde B_2]
             +2(c_2+c_3)c_5\cdot
[m_b(\epsilon_3+\epsilon_4)+2m_b\tilde\epsilon_2]
            {\Big \}}.
\end{eqnarray}

\subsection{The effective mass of the decaying heavy quark}

The masses of the acting heavy quarks 
in a decay must be treated carefully
although the bound-state effects
make the problem complicated and obscure. 
It is commonly accepted that 
if the charm quark appears as 
a decay product, the mass should be its running one at the
energy scale of the decaying quark or the meson, whereas, 
if it appears as the `parent(s)' of the decay,
the quark (antiquark) is not "free", but in a bound state,
thus the pole mass should be taken and the bound-state 
effects on the mass must be taken into
account too. Especially in spectator mechanism
the decay possibility of the heavy quark
is very sensitive to the value of its `adopted' mass,
hence what a value of the quark mass adopted in the
estimate must pay special care. Narison \cite{Narison} used 
the QCD sum rules to estimate
the mass difference $M_{b(c)}^{NR}-M_{b(c)}^{PT2}$ where $M^{PT2}$
is the short-distance perturbative pole mass and $M^{NR}$ is the
long-distance QCD-related effective mass up-to two-loops. Whereas
the authors of \cite{Kot} attributed such effects into a factor which
is multiplied to the decay width of the "free" quark.

Here instead of deriving the modification factor with
a relatively large uncertainty, we treat the problem
phenomenologically i.e. by introducing a parametrization
\begin{equation}
\label{del}
M_Q^{eff}=M_Q^{pole}-\Delta,
\end{equation}
where $\Delta$ manifests the bound-state effects, and 
it will be fixed phenomenologically. Note here that for each 
heavy meson there are three quantities: lifetime (total width), 
inclusive semileptonic branching ratio and pure leptonic branching
ratio which may be used for phenomenological analyses, so
the estimates here are still well-determined even when we 
introduce the parameter $\Delta$ here.

In the next section, we will discuss $\Delta$
and other related parameters more precisely.

With all the formulae derived above and the hadronic matrix elements, we
can make numerical evaluation of the lifetime of $B_c$ straightforwardly.

\section{Numerical results}

\indent

Since we carry out the estimate of the lifetime of $B_c$ with a 
`gloable' comparison to all of the heavy and double
heavy mesons, so the determination of all
of the parameters by fitting the existence experimental data is
`over-determined' for our goal and has certain level
tests. Therefore we evaluate the lifetimes, the semileptonic branching
ratios and the pure leptonic branching ratios for all the mesons
$D^\pm, D^0, D_s, B^\pm, B^0, B_s$ and $B_c$ in this section in turn
and present the numerical results in this section.

\subsection{For the heavy mesons $D$ and $B$}

To evaluate the lifetimes of $D^0, D^\pm, D_s, B^0, B^\pm, B_s$ mesons and
their branching ratios of the semileptonic decays, we use the formulae
given in sections 2.1, 2.2
and the appendix. The parameter values are taken as follows.
$|V_{cs}|=0.974, |V_{ud}|=0.975$, $\alpha_s(m_c)=0.29$,
$c_1(m	w_c)=1.30, c_2(m_c)=-0.57\cite{Cheng}$,
$B1=B2=1, \epsilon_1\simeq -0.05, \epsilon_2=0\cite{Cheng}$, the
decay constants of D mesons
$f_D=160$ MeV, $f_{D_s}=190$ MeV.
In the evaluation of the Pauli interference $PI$
contribution to $D$ decay width, we take the
$p_-^2=(p_c-p_{\bar q})^2$ value as $0.5$ $M_D^2$
as done in ref.\cite{Chernyak}.

We take $m_b^{pole}=5.02$ GeV, $m_c^{pole}=1.88$ 
GeV\cite{Yudurain97,Yudurain00}.
By eq.(\ref{run}), we have the running mass of the charm quark at
various energy scales as
$$\overline m_c(m_c)=1.67\;{\rm GeV},\;\;\;
\overline m_c(m_b)=1.41\;{\rm GeV},\;\;\;
\overline m_c(m_{B_c})=1.37\; {\rm GeV}.$$

By fitting data \cite{Cheng}, we should have
the quark masses as $m_s=125$ MeV,
$m^{eff}_c=1.65$ GeV respectively. Then we obtain the D meson
lifetimes:
$\tau(D^0)=0.419$ ps, $\tau(D^-)=1.06$ ps, $\tau(D^-_s)=0.446$ ps,
and the branch ratio of the semileptonic decay of $D^0$ meson
$B_{SL}(D^0)=6.9 \%$. Comparing to the experimental data:
$\tau(D^0)=0.415\pm 0.04$ ps; $\tau(D^\pm)=1.057\pm 0.015$ ps;
$\tau(D_s)=0.467\pm 0.017$ ps and $B_{SL}(D^0)=6.75\pm0.29\%$,
one can see the fit is quite well.

For the estimate of B meson lifetimes,
we take the  mass of the charm quark at final states $m_c$ to be
running mass, namely it is different from the pole mass of charm 
quark in $D$-decays but the running mass at the energy scale
$m_b$ i.e. $\bar {m_c}(m_b)=1.41$ GeV.
When calculating $PI$ contribution to the $B^-$ decay width,
we take the value of $p_-^2=(p_b-p_{\bar u})^2$ approximately to
be $0.8$ $M_B^2$ \cite{Chernyak}.
For the other parameters in the numerical computations the values
are adopted: 
$|V_{cb}|=0.04$, $\alpha_s(m_b)=0.20$,
$c_1(m_b)=1.150$, $c_2(m_b)=-0.313$ \cite{He},
$\alpha=1.06, \beta=1.32$ \cite{Yang},
$B1=B2=1$ and $\epsilon_1=-0.14, \epsilon_2=-0.08$ \cite{Cheng98}.
The decay constants: $f_{B}=200$ MeV and $f_{B_s}=220$ MeV.
Furthermore, taking the $b$ quark pole
mass $m_b^{pole}=5.02$ GeV and $m_b^{eff}=4.89\sim 4.91$ GeV,
we obtain the results that
$$\tau(B^0)=1.54{\rm ps},\;\;
\tau(B^-)=1.74{\rm ps},\;\;\tau(B^-_s)=1.56{\rm ps},\;\;
B_{sl}(B^0)=11.2\%, \;\; {\rm if}\; m_b^{eff}=4.89{\rm GeV}\; ;$$
$$\tau(B^0)=1.52{\rm ps},\;\;
\tau(B^-)=1.71{\rm ps},\;\;\tau(B^-_s)=1.54{\rm ps},\;\;
B_{sl}(B^0)=11.2\%, \;\; {\rm if}\; m_b^{eff}=4.90{\rm GeV}\; ;$$
$$\tau(B^0)=1.50{\rm ps},\;\;
\tau(B^-)=1.68{\rm ps},\;\;\tau(B^-_s)=1.51{\rm ps},\;\;
B_{sl}(B^0)=11.2\%, \;\; {\rm if}\; m_b^{eff}=4.91{\rm GeV}\; ,$$
where $B_{sl}$ indicates the branching ratio of the semileptonic
decay. Comparing with the experimental data 
$\tau(B^0)=1.56\pm 0.04$ ps, $\tau(B^\pm)=1.65\pm 0.04$ ps,
$\tau(B_s)=1.54\pm 0.07$ ps and $B_{SL}(B^0)=10.5\pm 0.008\%$
we can see the fit is quite good. Accouding to the definition
of $\Delta$, we have $\Delta_c\equiv 
m_c^{pole}-m_c^{eff}=0.23\;{\rm GeV}$
and $\Delta_b\equiv m_b^{pole}-m_b^{eff}=0.11\sim
0.13\;{\rm GeV},$ that is understandable. 

With all the parameters obtained by fitting the lifetimes of
$B^0, B^\pm, B_s,D^0, D^\pm, D_s$ and the branching ratios of the
semileptonic decays of B and D mesons, we are proceeding to evaluate the
lifetime of $B_c-$meson and its semileptonic decay rate.

\subsection{For the double heavy meson $B_c$}

The spectator componet contribution to the $B_c$ lifetime is a 
sum of the individual $\bar b$ and $c$ quark decays, while 
leaving the other one as a spectator. When evaluating
this contribution, $m_b$ is its pole value
at $p_b^2=m_b^2$, and $m_{\bar c}$ also the pole value
$m_c(m_c)$. Whereas for the non-spectator contributions, i.e. 
the $WA$ and $PI$ pieces, the corresponding energy scale for the running 
charm-quark mass in the final state is taken as $M_{B_c}$.
Now let us take the relevant parameters for $B_c$ as follows:
$M_{B_c}=6.25$ GeV, $M_{B_c}^*=6.33$ GeV, $B_1=B_2=1, 
\epsilon_1=-0.14, \epsilon_2=-0.08$.
For the decay constant, we adopt Eichten and Quigg's one
$f_{B_c}=500$ MeV\cite{Quigg} and the lattice one
$f_{B_c}=440$ MeV\cite{Davies} respectively.
Furthermore in the calculation of the $PI$ contribution, the quantity
$p_-^2 = (p_b-p_{\bar c})^2 \simeq 2m_b^2+2m_c^2-M_{B_c}^2$
is taken approximately. With the parameters described above, we
obtain the  numerical results and tabulate them in Table 1.

\vspace{2mm}
\begin{center}
{\bf Table 1: The results for $B_c$ meson}
\begin{tabular}{|c|c|c|c|c|c|c|c|c|}\hline
$f_{B_c}$         &$\tau_{B_c}$                   & $\Gamma^{pen.}$
                  & $\Gamma^{b\rightarrow c}$     & $\Gamma^{c\rightarrow s}$
                  & $\Gamma^{WA}$                 & $\Gamma^{PI}$
                  & $\Gamma(\tau\nu)$             & $B_{SL}$    \\ \hline
$440$MeV         &$0.362$ (ps)              & $3.4\%$
                 &$22.8\%$                  & $70.9\%$
                 &$13.4\%$                  & $-7.1\%$
                 &$0.078$ ps$^{-1}$         & $8.7\%$  \\  \hline
$500$MeV         &$0.357$ (ps)              &$4.3\%$
                 &$22.4\%$                  &$69.7\%$
                 &$16.9\%$                  &$-9.0\%$
                 &$0.100$ ps$^{-1}$         &$8.4\%$   \\   \hline
\end{tabular}
\end{center}

In the table $f_{B_c}$ denotes the decay constant; $\tau_{B_c}$: the
lifetime of $B_c$; $\Gamma^{pen.}$: the contribution from the interference
between the penguin and `tree' terms; $\Gamma(\tau\nu)$: the
width of the pure leptonic decay ($\tau$ chennal only but
almost equal to the total) and the $B_{SL}$: the
branching ratio of the semileptonic decay of the meson $B_c$.
\vspace{0.3cm}

Since both the `parents' $\bar b$ and $c$ quarks reside in a bound
state ($B_c$ meson), the problem how to choose the value
of the masses $m_b$ and $m_c$ emerge as in the cases
of the heavy mesons $D$ and $B$ etc, but when taking all the 
parameters fixed by fitting data as the above, we obtain the
results presented in table 1. 

Let us discuss the bound-state effects on the $\bar b$ and 
$c$ quark-masses in $B_c$ meson more precisely. Because $B_c$
includes two heavy quarks i.e. is a double heavy meson, 
the bound-state effects might be 
greater than in the heavy mesons $B,D$. 
We think that the values $m_c^{eff}$ and $m_b^{eff}$
might be smaller than $m_c^{eff}=1.65$ GeV and $m_b^{eff}=4.9$ 
GeV that we obtained in $B$ and $D$ decays. Phenomenologically, if
in $B_c$ meson, $m_c^{eff}(B_c)=1.55$ GeV, $m_b^{eff}(B_c)=4.85$
GeV, we obtain $\tau(B_c)\approx 0.47$ ps, which occationally is 
closer to the center value of the $B_c$ lifetime measured
recently\cite{Fermi}. In this case, $\Delta_c=0.33$ GeV and
$\Delta_b=0.17$ GeV. Because the rates
of direct $\bar b$ and $c$ decays, which dominate the lifetime of
$B_c$ meson, are proportional to $(M_Q^{eff})^5$,
the results are so sensitive to the effective
masses.

\section{Conclusion and Discussion}

\indent

In this work, we estimate the lifetime of $B_c$ 
in the `unique' theoretical framework where the 
nonspectator effects are taken into account properly
and the necessay parameters are determined
by fitting the data of the heavy mesons $B^0$, $B^\pm$,
$B_s$, $D^0$, $D^\pm$, $D_s$ on the lifetimes and the branching ratios 
of the inclusive semileptonic decays as input. 
 
Not only the uncertainties in the estimate are discussed
but also the physical parameters appearing in the
estimation are fixed in reasonable regions. 

Even though not all of the parameters are fixed by fitting
the avialable data for the heavy mesons $B^0$, $B^\pm$, $B_s$, $D^0$,
$D^\pm$, $D_s$, in order to carry out the estimate 
we make some reasonable assumptions or conjections.
In fact, in terms of the lattice calculation, QCD sum rules 
and other approaches. the parameters, 
such as $B_1$ and $B_2$ (the factors in the hadronic matrix 
elements, the values manifest the deviation to the vacuum 
saturation), may be determined at certain accurate level. 
The deviation from model-dependent calculation 
is small, and can be neglected in practice.
Some of the parameters, such as $\epsilon_1$ and $\epsilon_2$, being
realized to relate to the non-factorization effects\cite{Buras},
may be calculated at the $D$-meson scale and the B-meson scale 
respectively in terms of the QCD sum
rules\cite{Cheng,Yang,Cheng98}. It is known that the 
numerical values obtained by the QCD sum rules may have 
errors about $10\sim 15\%$, but they still can be
used as inputs to the phenomenological calculations without
causing too large errors. In this work we also carefully
consider the quark masses and take into account of the
bound-state effects. The consistency of our numerical results of
B and D mesons confirms the validity of the parameter regions.

The earlier estimations on the lifetimes of B and D mesons and the
semileptonic decay rates obviously deviate from the data. Luke,
Savage and Wise\cite{Luke} pointed out that in decay
$c\rightarrow X\bar e\nu_e$, the contribution of $\alpha_s^2$ order
is of the same magnitude as that of ${\cal O}(\alpha_s)$ and
this higher order correction suppresses the semileptonic decay rate
of D-meson. Taking into account this fact, we obtain numerical
results of lifetimes of D-mesons and their semileptonic decay rate,
and find that they are satisfactorily consistent with data. Whereas, 
for B-meson decays, the $\alpha_s^2$ order correction, as well
as the ${\cal O}(\alpha_s)$ correction, are smaller. With these 
corrections concerned, the results for B-mesons are also
consistent with data within the experiment tolerance region. 
All these imply that the parameters taken as the above are reasonable.

When evaluating the $B_c$ lifetime and its inclusive semileptonic 
decay rate, some new aspects must be taken into account. First there 
are several new operators in the effective Lagrangian playing roles. 
Their appearance is due to non-negligible $m_c$, whereas in B and D cases, 
the light quark mass $m_q$ is ignored with quite high accuracy.
Correspondingly, several new hadronic matrix elements are induced by these
operators. Some of them are also proportional to $B_1$ and
$B_2$, which exist in the expressions for B and D meson decays, as
long as the factorization theorem and the vacuum saturation hold.
But in the non-factorization contributions, new parameters appear and
$\epsilon_3\sim \epsilon_6$ is assumed. In this work, we have
taken a naive symmetry consideration to let $\epsilon_{3,4}\simeq
\epsilon_2$ and $\epsilon_{5,6}\simeq \epsilon_1$.

As discussed in the introduction, in the $B_c$ case, the 
interference between the penguin and tree terms is not negligible.
Namely, the penguin contribution to $B_c$ lifetime is much more
important than to B and D decays. Our results confirm this
allegation and we have found the contribution from the interference
can be as large as $3\sim 4\%$ of the total width.
This fraction is measurable in accurate measurements. Since
direct measurement of penguin diagrams would be interesting,
these sizable value can be encouraging for future experiments.

The lifetime of $B_c$-meson is estimated as $0.37\sim 0.367$ ps for
$f_{B_c}\sim 440\sim 500$ MeV, which are smaller than the central 
value of the measurement $\tau_{B_c}=0.46^{+0.18}_{-0.16}$(stat)$\pm
0.063$(syst) ps\cite{Fermi}. Our earlier estimation of 
$\tau_{B_c}$\cite{Chang} was 0.4 ps. Anisimov {\it et al.} estimated
$\tau_{B_c}$ in the light-front constituent quark model and obtained
$\tau_{B_c}=0.59\pm 0.06$ ps, which is larger than the measured
value. In our estimation, we use the values of $M_{B_c}$ and
$M_{B_c}^*$ as $6.25$ GeV and $6.33$ GeV \cite{Bigi}, whereas the
measurement is $M_{B_c}\sim 6.40\pm 0.39$(stat)$\pm 0.13$(syst) GeV.
When the bound-state effects on $\bar b$ and $c$ masses are
reasonably taken into account, we can have $\tau_{B_c}\sim 0.47$ ps,
which is very close to the present experimental center value 
of the $B_c$-meson lifetime. As noted, the change of $f_{B_c}$
itself only does not influence the result much, e.g. as $f_{B_c}$
changes from $440$ MeV to 500 MeV, $\tau_{B_c}$ varies $1\%$ only.
Whereas, a large error in mass of the decay parents would 
result in an estimate error of about $4\%$, therefore, the more
accurate lifetime and mass of $B_c$ meson will test the framework
adopted here the deeper.

The more accurate experimental measurements will shed 
fresh lights on the framework adopted here, which
involving the effective heavy flavour theory and
the duality between quark states and hadronic states
$$\sum\limits_{i,j} |q_i,g_j\rangle \langle q_i,g_j| = 
\sum\limits_{k} |h_k\rangle \langle h_k|$$ 
etc. It will help to clarify many uncertainties.

\vspace*{0.6cm}
\noindent
{\Large\bf Acknowledgements}

This work was supported in part by the National Natural Science Foundation
of China. One of the author (C.-H. Chang) would like to thank TH.-Division
of CERN for their warm hospitality, since this paper is complished
its final composition during his short visit of CERN.

\vspace{1cm}

\noindent{\bf Appendix A}

The semileptonic and non-leptonic decay rates of $b$ quark through order
$1/m_Q^2$ are given as following \cite{Bigi,Cheng97}.
\begin{eqnarray}
\nonumber
 \Gamma_{SL}(H_b) & = &
  \Gamma_0^{(b)} \cdot
    \eta(x_c,x_l,0)\cdot
    {\Big [}
    I_0(x_c,0,0)\langle H_b|\bar bb|H_b\rangle -
 \frac{2\langle \mu_G^2\rangle _{H_b}}{m_b^2}
 I_1(x_c,0,0)
 {\Big ]}\;;
 \\
 \nonumber
\Gamma_{NL}(H_b) & = &
        \Gamma_0^{(b)}\cdot N\cdot
      {\Big \{}
      (c_1^2+c_2^2+\frac{2c_1c_2}{N})\cdot [(\alpha I_0(x_c,0,0)
      +\beta I_0(x_c,x_c,0))\langle H_b|\bar bb|H_b\rangle
   \\         \nonumber
   & &
    -\frac{2\langle \mu_G^2\rangle _{H_b}}{m_b^2}
    (I_1(x_c,0,0)+I_1(x_c,x_c,0))]
    \\
   & &
     -8\frac{\langle \mu_G^2\rangle_{H_b}}{m_b^2}\frac{2c_1c_2}{N}\cdot
    [I_2(x_c,0,0)+I_2(x_c,x_c,0)]
     {\Big \}}\;,
\end{eqnarray}
where
\begin{equation}
\Gamma_0^{(b)} \equiv
\frac{G_F^2m_b^5}{192\pi^3}|V_{cb}|^2;
\end{equation}
and the following notation has been used:
$I_0$, $I_1$ and $I_2$ are phase-space factors, namely
\begin{eqnarray}
   \nonumber
   I_0(x,0,0)&=&(1-x^2)(1-8x+x^2)-12x^2\log x,
\\      \nonumber
   I_1(x,0,0)&=&\frac{1}{2}(2-x\frac{d}{dx})I_0(x,0,0),
\\       \nonumber
   I_2(x,0,0)&=&(1-x)^3,
\\       \nonumber
   I_0(x,x,0)&=&v(1-14x-2x^2-12x^3) + 24x^2(1-x^2)\log \frac{1+v}{1-v},
\\        \nonumber
   I_1(x,x,0)&=& \frac{1}{2}(2-x\frac{d}{dx})I_0(x,x,0),
\\     \nonumber
  I_2(x,x,0)&=&v(1+\frac{x}{2}+3x^2)-3x(1-2x^2)\log
  \frac{1+v}{1-v},
\\
  x_c&=&(\bar m_c/m_b)^2,  \hspace*{0.5cm}  v=\sqrt{1-4x}\;,
\end{eqnarray}
with $I_{0,1,2}(x,x,0)$ describing the $b\rightarrow c\bar cs$
transitions.

And for $\eta(x_c,x_l,0)$, which is the QCD radiative correction
to the semileptonic decay rate. Its general analytic expression is
given in \cite{Hokim}. The special case $\eta(x,0,0)$ is given in
\cite{Nir} and it can be approximated numerically by \cite{Kim,Cheng97}
\begin{equation}
  \eta(x,0,0)\cong 1-\frac{2\alpha_s}{3\pi}
  {\Big [}
    (\pi^2-\frac{31}{4})(1-\sqrt{x})^2+\frac{3}{2}
  {\Big ]}.
\end{equation}
For the decay $b\rightarrow c\tau\nu$, according to
\cite{Bagan} we roughly have
\begin{equation}
\Gamma(b\rightarrow c\tau\nu)\sim 0.25\Gamma(b\rightarrow ce\nu).
\end{equation}
The expressions are simpler for $c\rightarrow s$:
\begin{eqnarray}
    \Gamma_{SL}(H_c) & = &
     \Gamma_0^{(c)} \cdot
      \eta(x_s,x_l,0)
   {\Big [}
      I_0(x_s,0,0)\langle H_c|\bar cc|H_c\rangle
     -\frac{2\langle \mu_G^2\rangle _{H_c}}{m_c^2}
    I_1(x_s,0,0)
    {\Big ]},
    \\
    \nonumber
   \Gamma_{NL}(H_c)& = &
      \Gamma_0^{(c)}\cdot N\cdot
           {\Big \{}
         (c_1^2+c_2^2+\frac{2c_1c_2}{N})\times
         [\alpha I_0(x_s,0,0)\langle H_c|\bar cc|H_c\rangle
      \\
      & &
        -\frac{2\langle \mu_G^2\rangle _{H_c}}{m_c^2}
          I_1(x_s,0,0)]
       -8\frac{\langle \mu_G^2\rangle_{H_c}}{m_c^2}\frac{2c_1c_2}{N}
        \cdot I_2(x_s,0,0)
         {\Big \}}.
\end{eqnarray}
where
\begin{equation}
  \Gamma_0^{(c)}\equiv \frac{G_F^2m_c^5}{192\pi^3}|V_{cs}|^2,
 \hspace*{0.3cm}    x_s=\frac{\bar m_s^2}{m_c^2} .
\end{equation}
and for the correction $\eta(x_s,x_l,0)$ in the c-decay case, we adopt
a numerical expression from \cite{Luke}. It reads
\begin{equation}
\eta_{SL}=1-2.08{\Big (}\frac{\alpha_s(m_c)}{\pi}{\Big )}
        -22.7{\Big (}\frac{\alpha_s(m_c)}{\pi}{\Big )}^2.
\end{equation}
For the dimension-three operator $\bar QQ$, the expectation value
can be expressed at follows:
\begin{equation}
\langle H_Q|\bar QQ|H_Q\rangle=1-\frac{\langle({\bf p}_Q)^2\rangle _{H_Q}}{2m_Q^2}
                +\frac{\langle \mu^2_G \rangle _{H_Q}}{2m_Q^2}
                +{\cal O}(1/m_Q^3);
\end{equation}
where $ \langle ({\bf p}_Q)^2\rangle\equiv\langle H_Q|\bar Q(iD)^2Q|H_Q\rangle $
denotes the average kinetic energy of the quark Q moving inside
the hadron and
$ \langle \mu^2_G\rangle_{H_Q}\equiv
\langle H_Q|\bar Q\frac{i}{2} \sigma\cdot GQ|H_Q\rangle $.

Based on refs.\cite{Bigi,Bigi1} the kinetic terms take the values
respectively as follows:
\begin{equation}
  \frac{\langle({\bf p}_b)^2\rangle _{B}}{m_b^2}\simeq 0.016\;,
  \frac{\langle({\bf p}_c)^2\rangle _{D}}{m_c^2}\simeq 0.21\;;
  \frac{\langle({\bf p}_b)^2\rangle _{B_c}}{m_b^2}\simeq 0.04\;,
  \frac{\langle({\bf p}_c)^2\rangle _{B_c}}{m_c^2}\simeq 0.4\;.
\end{equation}
For the chromomagnetic operator one finds
$ \langle \mu_G^2\rangle_{P_Q}\simeq
\frac{3}{2}m_Q (M_{V_Q}-M_{P_Q})$,
where $P_Q$ and $V_Q$ denote the pseudoscalar and vector mesons,
respectively.

\end{document}